
\input PHYZZX
\PHYSREV
\date={June 1992}
\Pubnum={\caps UPR-514-T
 }
\def\ghat#1#2{{\hat{g}_{#1}^{#2}}}

\def\etal{{\it et al.}}
\def\to{\rightarrow}
\titlepage
\title{$V'Z$ and $V'W$
 Production as  Tests of Heavy Gauge Boson
Couplings at Future Hadron Colliders
}
\frontpageskip=0.5\medskipamount plus 0.5 fil
\author{Mirjam Cveti\v c and  Paul Langacker}
\address{Department of Physics\break
University of Pennsylvania\break
Philadelphia, PA 19104--6396\break}

\abstract{
We point out that the
production cross section of $pp\rightarrow V'V$, with
$V'=W',Z'$ and
$V=W,Z$ is a useful diagnostic  of $V'$ gauge couplings
at future hadron colliders. For $M_{Z'}\simeq1$ TeV it would allow
determination of combinations of $Z'$
gauge couplings to the quarks to around 10    percent.
An analysis of the extraction of gauge couplings from the complementary
tests: forward-backward asymmetry,  rare decays
$pp\rightarrow V'\rightarrow f_1{\bar f_2}V$, and the production
cross section $pp\rightarrow V'V$ is given in a model-independent
framework. Four ratios of charges are needed to characterize a general
gauge theory with an additional family-independent $U_1'$ factor.  We
show that there are four
 functions of these ratios
observable at hadron colliders, but for projected SSC and LHC
luminosities only two combinations can be extracted.  These yield a
significant
 discrimination between interesting
GUT-motivated models.  Clean tests of whether a new $W'$ couples to
right-handed currents, of the ratio $g_R/g_L$ of gauge couplings, and
of the non-abelian vertex in left-right  symmetric models are
described.
}

\line{PACS \# 12.15, 12.10, 11.15\hfill}
\endpage

\chapter{Introduction}
Extended gauge structures are an essential part of grand unification.
In string theory the gauge
group at the compactification scale is generally larger than that of
the standard model.  Although there is no a priori reason (except in
some cases of restrictive particle representation) to constrain the mass
of the new gauge boson to the accessible range of a few TeV,
the existence of new
gauge bosons in this energy range is a plausible consequence of a
number of types of new physics\rlap.\Ref\rfi{M.~Cveti\v c
and P.~Langacker, Phys. Rev. D{\bf 42}, 1797 (1990).}

Current limits on extra neutral gauge bosons ($Z'$) are relatively weak.
The analysis of $Z$-pole, weak neutral current, and collider data puts
lower bounds\REF\rfii{
G.~Altarelli, {\it et al.}, Phys. Lett. {\bf B263},
459 (1991), {\it ibid.} {\bf B261}, 146 (1991);
J.~Layssac, F.~M.~Renard, and C.~Verzegnassi, Z. Phys. {\bf C53},
97 (1992); F.~M.~Renard and C.~Verzegnassi, Phys. Lett.
{\bf B260}, 225 (1991).}\REF\rfiii{
 K.~T.~Mahanthappa and P.~K.~Mohapatra, Phys. Rev. D
{\bf 43}, 3093 (1991); P.~Langacker, Phys. Lett. {\bf B256}, 277 (1991).}
\REF\delA{F.~del Aguila, W.~Hollik
 J.~M.~Moreno, and M.~Quiros, Nucl. Phys. {B372}, 3 (1992);
 F.~del Aguila, J.~M.~Moreno, and M.~Quiros,
Nucl. Phys. {\bf B361}, 45 (1991); Phys. Lett. {\bf B254},
 479 (1991);   M.~C.~Gonzalez-Garcia and
J.~W.~F.~Valle, Phys. Lett. {\bf B259}, 365 (1991).}\REF\rfv{
P.~Langacker and M.~Luo, Phys. Rev. {\bf D44}, 817 (1991);
U.~Amaldi, \etal, Phys. Rev. {\bf D36}, 1385 (1987); S.~Durkin and
P.~Langacker, Phys. Lett. {\bf166B}, 436 (1986).
The  sensitivity of future precision experiments is discussed in
P.~Langacker, M.~Luo and A.~K.~Mann, Rev. Mod. Phys. {\bf64}, 87
(1992).}
\refmark{\rfii,\rfiii,\delA,\rfv}\ on the masses of various types
of $Z'$'s in the range of 160--400 GeV, with stronger limits around
500--1000 GeV in some specific models in which the mass and the $Z-Z'$
mixing angle are related.

The limits on the new charged gauge bosons
$W^{\prime\pm}$ are
more constrained in specific models, in particular, in
the left-right symmetric models\Ref\rfip{
J.~C.~Pati and A.~Salam, Phys.
Rev. Lett. {\bf 31}, 661 (1973); Phys. Rev. {\bf D10}, 275 (1974);
R. Mohapatra and J. C. Pati, Phys. Rev. {\bf D11}, 566 (1975).}
based on the extended gauge structure $SU_{2L}\times SU_{2R}\times
U_{1(B-L)}$.
  For the models with $g_L=g_R$
(respective gauge couplings for $SU_{2L,2R}$) and equal magnitude of the
left-handed and right-handed quark-mixing matrix elements, the bound on
the mass of the heavy charged $W'$ is $M_{W'}>1.4\hbox{~TeV}$, based on
the $K_S-K_L$ mass difference\rlap,\Ref\rfiip{
G. Beall, M. Bander,
and A. Soni, Phys. Rev. Lett. {\bf 48}, 848 (1982).}
and the $W-W'$ mixing angle is $|\theta_+|<0.003$, from
universality\rlap.\Ref\rfiiip{
L. Wolfenstein, Phys. Rev. {\bf D29}, 2130 (1984).}
For general left-right symmetric models these bounds are much
weaker\rlap:\Ref\rfivp{P. Langacker
and U. Sankar, Phys. Rev. {\bf D40}, 1569 (1989).}
$g_L M_{W'}/g_R>300\hbox{~GeV}$ and $g_R|\theta_+|/g_L<0.013$.
Stronger limits follow from CP violation unless there is fine tuning.%
\Ref\rfLoWy{D.~London and D.~Wyler, Phys. Lett. {\bf B232}, 503
(1989).}

\REF\LRR{ P.~Langacker, R.~Robinett, and J.~Rosner, Phys. Rev.
 {\bf D30}, 1470 (1984).}
\REF\Bar{ V.~Barger, {\it et al.}, Phys. Rev.   {\bf D35}, 2893
(1987).}
\REF\DuLa{ L.~Durkin and P.~Langacker, Phys. Lett. {\bf B166},
436 (1986); F.~del~Aguila, M.~Quiros, and F.~Zwirner, Nucl. Phys.
{\bf B287}, 419 (1987); {\bf284}, 530 (1987); J.~Hewett and T.~Rizzo in
{\sl Proceedings of the 1988 Snowmass Summer Study on High Energy
Physics in the 1990's}, Snowmass, CO 1988; P.~Chiappetta  {\it et
al.}, in the {\sl Proceedings of the Large Hadron Collider Workshop},
Aachen, Germany, 1990.}
\REF\HRII{
  J. ~Hewett and  T. ~Rizzo, Phys. Rev. {\bf D45}, 161 (1992).}
On the other hand, the
$Z'$ can be produced\refmark{\LRR,\delA} (and
clearly detected via leptonic decays)
at the LHC and SSC if its mass does not exceed around
5 TeV\rlap.\refmark{\LRR -\HRII}

\REF\NLC{The possibilities at a 500 GeV $e^+ e^-$ collider have been
studied by A. Djouadi {\it et al.,} DESY preprint, 1992.}
The immediate goal after the discovery of a new gauge boson would be to
try to understand its origin and properties.\refmark{\NLC}
Subsequent tests should be able to address and separate
the following:

\item{(i)} The nature of the gauge couplings of new gauge bosons to
ordinary fermions, including the pattern of charges for a new $Z'$,
whether a new $W'$ couples to $V+A$ currents, and the overall strength
of the gauge coupling.

\item{(ii)} The nature of the symmetry breaking structure.

\item{(iii)} The coupling of the gauge bosons to exotic fermions
and supersymmetric partners.

Although the three phenomena are interconnected, the aim is to
study experimental signals that can separate them.
Some aspects of the nature of symmetry breaking structure,
in particular, the decays $V'\rightarrow VV$
($V=W,Z$; $V'=W',Z'$))
\REF\rfx{
F.~del Aguila, M.~Quiros, and F.~Zwirner, Nucl. Phys.
B {\bf284}, 530 (1987); P.~Kalyniak and M.~Sundaresan, Phys. Rev. D
{\bf35}, 75 (1987).}\REF\rfxi{
N.~Deshpande, J.~Gunion, and F.~Zwirner, in the
{\sl Proceedings of the Workshop on Experiments Detectors and
Experimental Areas for the Supercollider} (Berkeley 1987); N.
Deshpande, J.~Grifols, and A.~M\'endez, Phys. Lett. {\bf B208}, 141
(1988).}\REF\rfxii{
F.~del Aguila, L.~Ammetller, R.~Field, and
L.~Garrido, Phys. Lett. {\bf B201}, 375 (1988); Phys. Lett. {\bf
B221}, 408 (1989).}
\REF\rfKaYu{G.~Kane and C.~P.~Yuan, Phys. Rev. {\bf D40}, 2231 (1989)}
\REF\rfix{M.~Cveti\v c, B.~Kayser, and P.~Langacker,
Phys. Rev. Lett. {\bf68}, 2871 (1992).}
\refmark{\rfx -\rfKaYu}   and the $M_{W'}/M_{Z'}$ mass
ratio\rlap,\refmark{\rfi,\rfix}
as well as
 \REF\rfxiii{M.~J.~Duncan and P.~Langacker, Nucl. Phys. {\bf B277},
285 (1986).}
the study of decays of $V'$ into exotic fermions\refmark{\rfxiii}
have been  studied earlier.

 In this paper we address the
 diagnostic study of the gauge couplings. One immediate
possibility is the measurement of $\sigma (pp\to V')
 B$ where $\sigma(pp\to V')
 $
is the total  production cross section   and
$B$ is the branching ratio into leptons.
$\sigma$ could be calculated for a given set of
$(Z',W^{\prime\pm})$ couplings to within a few percent. However,
the theoretical
branching ratio, $B\equiv\Gamma(Z'\to\ell^+\ell^-)/\Gamma_{tot}$,
where $\ell=(e,\mu,\hbox{~or both})$ and $\Gamma_{tot}$ is the total
width, with an analogous definition for $W$,
is model
dependent because it depends on the contribution of exotic
fermions and supersymmetric partners to the $V'$ width. These could
easily change $\sigma B$ by a factor of 2.  Thus $\sigma B$ cannot be
useful as a diagnostic test for the $V'$ gauge coupling; however, it
would be a useful indirect probe for the existence of the exotic
fermions or superpartners.

On the other hand, from measurements of the total width\refmark\HRII
$\Gamma_{tot}$,  which could be
determined from the line shape
of $Z'\rightarrow \ell^+ \ell^-$, and $\sigma B$ one obtains
$\sigma \Gamma(Z'\rightarrow \ell^+\ell^-)\equiv\sigma B\Gamma_{tot}$.
This  probes the absolute magnitude of the gauge couplings
in the combination $g^4_2 [a(\ghat{L2}{u~2}
+\ghat{R2}{u~2} ) + b   (\ghat{L2}{d~2} +\ghat{R2}{d~2})]
(\ghat{L2}{\ell~2}+\ghat{R2}{\ell~2})$,
where $\hat{g}_{L2}^f$ ($\hat{g}_{R2}^f$) is the charge of the left
(right) -handed fermion $f$, $g_2$ is the gauge coupling of the new
boson, and $ a    $ and $ b   $ are calculable coefficients that depend
on the $V'$ mass, the $CM$ energy, and the proton structure functions.
The charges are characteristic of the gauge group.  Once this is
identified (as described below), $\sigma B\Gamma_{tot}$ would determine
the gauge coupling $g_2$, which is a useful probe of the
pattern of  spontaneous symmetry breaking in grand unified theories%
\Ref\RRref{R.~Robinett, {\sl Phys. Rev.} {\bf D26}, 2388 (1982);
  R.~Robinett and J.~Rosner, {\sl Phys. Rev.} {\bf D25}, 3036 (1982) and
 {\bf D26}, 2396 (1982).} and left-right symmetric models.

The production cross section for $pp\rightarrow V'g$, with $g$ a
gluon, has a large rate but
is not a useful probe. It has the same branching ratio uncertainties as
$\sigma B$.  Furthermore, since
gluons couple to the vector component of fermionic
currents, this process measures the same combination
of gauge couplings as $\sigma(pp\rightarrow V')$, so that the ratio of
cross sections yields no new information.

\REF\footI{Note, that in Ref. \rfxiv\ it was erroneously
mentioned that such a cross section would be at
most marginal at LHC. }
\REF\rfxiv{M. Cveti\v c and P. Langacker,
Univ. of Pennsylvania Preprint UPR--487--T (September 1991),
to appear in Phys. Rev. D.}
In this paper we propose\refmark{\footI}
 the cross section $pp\rightarrow V' V$ as a useful
probe\rlap,\Ref\rffti{As shown in J. Hewett and T. Rizzo, Phys. Rep.
{\bf 183}, 193 (1989),
the process $pp\rightarrow V'V'$ is too small to
be useful in reasonable models  because of a kinematic suppression
and also smaller couplings compared to $pp\rightarrow V'V.$} since it
measures a different combination of couplings:
$g_1^2g_2^2(\hat{g}_{L1}^{q~2}\hat{g}_{L2}^{q~2}
 + \hat{g}_{R1}^{q~2}\hat{g}_{R2}^{q~2})$.
Here $g_1$ and $g_2$ are the gauge coupling constants of the $V$ and
$V'$, respectively, while $\hat{g}^q_{(L,R)1}$ and
$\hat{g}_{(L,R)2}^{q}$ are the corresponding charges of the
(left, right) handed quarks.
In this case the uncertainty due to
 the exotics and $g_2$ is  removed by dividing
the   production rate by that for $pp\rightarrow V'$.
For leptonic decays of $Z'\rightarrow e^+e^- (\mu^+\mu^-)$ and
$(Z,W)$ decaying into either leptons or  quarks
this process is
virtually free of backgrounds.
The rate is sufficient to give good statistics for
$M_{V'}\sim1\hbox{~TeV}$ and provides a diagnostic test for the gauge
coupling of $V'$ to quarks.

\REF\rfxiiip{T. Rizzo, Phys. Lett. {\bf B192}, 125 (1987).}
This is complementary to the rare
decays\refmark{\rfxiiip}
\REF\PACO{F. del Aguila, B. Alles, Ll. Ametller and
A. Grau, University of Granada preprint, UG-FT-22/92 (June 1992).}
\REF\HR{J. Hewett and T. Rizzo, Argonne National Laboratory
preprint, ANL-HEP-PR-92-33 (June 1992).}
 recently  proposed \refmark{
\rfxiv} and studied in detail\refmark{\rfxiv, \PACO , \HR}
 as a useful diagnostic of
the heavy gauge boson couplings to  the ordinary fermions.
Such
rare decays involve $V'\rightarrow f_1\overline{f_2} V$,
where $f_{1,2}$ are
ordinary fermions.  For $V=Z$ the combination
$g_1^2g_2^2(\hat{g}_{L1}^2 \hat{g}_{L2}^2 +
\hat{g}_{R1}^2\hat{g}_{R2}^2)$ is measured, while for $V=W$,
such decays project out $\hat{g}_{L2}$ couplings.  Although
the rates are suppressed by a factor of $\alpha/2\pi$ compared with
$V'\rightarrow f_1\overline{f_2}$, they have a logarithmic enhancement
proportional to $\log^2(M^2_{V'}/M^2_V)$, closely related to collinear
and infrared singularities of QED\rlap.\REF\rfxv{
 T.~Kinoshita, J. Math. Phys. {\bf 3}, 650 (1962);
T. ~D. ~Lee and ~M. ~Nauenberg, Phys. Rev. {\bf 133}, B1549
(1964).}\REF\rfxvi{
W. Marciano and D. Wyler, Z. Phys.
{\bf C3}, 181 (1979).}\refmark{\rfxv,\rfxvi}
The gold-plated events (free of major backgrounds)
turn out to be\refmark\rfxiv
 $Z'\rightarrow W e {\bar\nu_e}(\mu{\bar\nu_\mu})$. Dividing
by the $Z'\rightarrow\ell^+\ell^-$ rate one has a clean
test of the coupling of $Z'$ to leptons.
In addition,
 the absence of $W^{\pm\prime}\rightarrow W^\pm e^+e^-(
\mu^+\mu^-)$ is an excellent test of the right-handedness of the
$W'$.

The third clean probe for gauge couplings is the forward-backward
asymmetry $A_{FB}$\refmark\LRR\ for the process $pp\rightarrow
Z'\rightarrow e^+e^-$ or $\mu^+\mu^-$. It can distinguish between
different models for $M_{Z'}$ up to few TeV, and tests a combination of
the couplings of $Z'$ to quarks and leptons.

This paper is organized as follows.
In Chapter 2 the formalism and  the models used in the calculation
are described.
The major part of the paper, Chapter 3,
involves a study of the proposed
production cross sections $pp\rightarrow V'V$, with
definitions of  appropriate ratios and backgrounds.
In Chapter 4  we  discuss the
extraction of detailed information about the gauge couplings.
We  point out how to study  family universality, the nature of
the enhanced gauge symmetry, and the values of the gauge
couplings for particular fermions based on the above
three diagnostic  probes. We show that considerable diagnostic
information can be obtained in hadron colliders.
Conclusions are given in Chapter 5.

\chapter{ Formalism}
The neutral current gauge interaction term in the presence of an
additional $U_1$ can be written as:
$$-L_{NC}=eJ_{em}^\mu A_\mu +g_1 J_1^\mu Z_{1\mu}+ g_2J_2^\mu Z_{2\mu},
\eqn\EQNC$$
with $Z_1$ being the $SU_2 \times U_1$ boson and $Z_2$ the additional
boson in the weak eigenstate basis.  Here
$g_1\equiv\sqrt{g_L^2+g_Y^2}=g_L/\cos\theta_W$, where
 $g_L$, $g_Y$ are the
gauge couplings of $SU_{2L}$ and $U_{1Y}$,
and $g_2$ is the gauge coupling of $Z_2$.  The GUT-motivated cases have
$g_2=\sqrt{5/3}\sin\theta_W g_1\lambda_g^{1/2}$, where $\lambda_g$
depends on the symmetry breaking pattern\rlap.\refmark\RRref\
If the GUT group breaks
directly to $SU_3\times SU_2\times U_1\times U_1'$, then $\lambda_g=1$.

The currents in \EQNC\  are:
$$J_j^\mu={1\over2}\sum_i{\bar\psi_i}\gamma^\mu
\left[\hat{g}^i_{Vj}-\hat{g}^i_{Aj}\gamma_5\right] \psi_i,
\quad j=1,2,
\eqn\CURR$$
where the sum runs over fermions, and the $\hat{g}^i_{(V,A)j}$
correspond to the vector and axial vector couplings of $Z_j$ to the
$i^{th}$ flavor.  Analogously, $\hat{g}^i_{(L,R)j}={1\over2}(
\hat{g}^i_{Vj}\pm \hat{g}^i_{Aj})$,
so that $\ghat{L1}{i} = t_{3L}^i - \sin^2 \theta_W q^i$ and
$\ghat{R1}{i} = - \sin^2 \theta_W q^i$, where $t_{3L}^i$
and $q^i$ are respectively the third component of weak isospin and
electric charge of fermion $i$.
We will illustrate
our study of $Z_2$ currents
with the following GUT, left-right (LR), and
superstring-motivated models:

\item{(i)}$Z_\chi$ occurs in $SO_{10}\rightarrow SU_5\times U_{1\chi}$,

\item{(ii)}$Z_\psi$ occurs in $E_6\rightarrow SO_{10}\times U_{1\psi}$,

\item{(iii)}$Z_\eta=\sqrt{3/8}Z_\chi-\sqrt{5/8}Z_\psi$ occurs in
superstring-inspired models in which $E_6$ breaks directly to a rank 5
group,

\item{(iv)}The general $E_6$ boson $Z(\beta)=\cos\beta
Z_\chi+\sin\beta Z_\psi$, where $0\le\beta<\pi$ is a mixing angle.
The $Z_\chi$, $Z_\psi$, and $Z_\eta$ are special cases with
$\beta=0,{\pi\over2}$ and $Z_\eta=-Z$
($\beta=\pi-\arctan\sqrt{5/3}$), respectively.

\item{(v)}$Z_{LR}$ occurs in left-right symmetric models,
where the ratio
$\kappa=g_R/g_L$ of the gauge couplings $g_{L,R}$ for $SU_{2L,2R}$,
respectively, parametrizes the whole class of models.
In this case $\lambda_g = 1$ by construction and $\kappa > 0.55$
for consistency\refmark{\rfix}.

\item{(vi)}$Z''$ has the same couplings as the ordinary $Z$; it cannot
occur in extended gauge theories, but could occur in composite models.
It is   a useful reference point for  comparing the sensitivity of
experimental signals. Note, however, that the more realistic cases {\it
(i)-(v)} have weaker couplings to the ordinary fermions.

The chiral couplings $\hat{g}_{(L,R)}$ for specific models are given in
Ref.~\rfv, but are repeated for convenience in Table~1.

\chapter{ $VV'$ Production Cross Sections}
Heavy neutral gauge bosons $Z'$ can be produced at future hadron
colliders (SSC and LHC) and can be  detected via the resultant leptonic
decays
$pp\rightarrow Z'\rightarrow e^+e^- (\mu^+\mu^-)$, and
the charged gauge bosons $W^{\pm\prime}$ are detected via the modes%
\Ref\rfval{If the $n_\ell$ is  heavy, its
subsequent decays may be observable and yield additional information.
M.~Cveti\v c, B.~Kayser, and P.~Langacker, unpublished.}\
$pp\rightarrow W^{\pm\prime}\rightarrow \bar \ell n_\ell(
\ell\bar  n_\ell)$, where $n_\ell$ is a  right-handed neutrino.
For given $Z'$ couplings the total cross
section $\sigma(pp\rightarrow Z')$ can be computed quite accurately,
since the quark cross section and distribution functions are known up to
${\cal O}(\alpha_s)$.  The cross sections are given in Refs.~\LRR-\HRII.
In the following we assume (for illustration) leptonic branching ratios
 $B$ corresponding to   decays
into {\bf16}-plets for $Z_\chi$ and $Z_{LR}$, {\bf27}-plets for $Z_\psi$
and $Z_\eta$, and {\bf15}-plets for $Z''$, and no superpartners.

Neutral
weak
eigenstates $Z_1$ and $Z_2$ can
 be identified with the mass eigenstates $Z$ and $Z^\prime$,
 respectively, because the $Z-Z'$  mixing is negligible,
as suggested from gauge theories as well as experiment.\refmark{
\rfii-\rfv}
The squared
amplitude (see Fig. 1a and Fig. 1b for the Feynman diagrams)
for the quark process $q{\bar q}\rightarrow Z'Z$ averaged
(summed)
over initial
(final) polarizations
can be written as:\Ref\BSM{
R. Brown, D. Sahdev
and K. Mikaelian, Phys. Rev. {\bf D20}, 1164 (1979);
R. Brown and K. Mikaelian, Phys. Rev. {\bf D19}, 992 (1979).
They also consider
the related process $q_1{\bar q_2}\rightarrow ZW$ in the standard
model.}
$$ {\cal M}_{Z,Z'}=
16\pi{{d\sigma}\over{d\hat{t}}}
=2{C_{ffZ}\over
\hat{s}^2}\left\{ \left( {\tilde{t}\over\tilde{u}}+
{\tilde{u}\over\tilde{t}}
\right) +2{{m_+}\over{\tilde{u}\tilde{t}}} -m_1m_2 \left(
{1\over\tilde{u}^2}+{1\over\tilde{t}^2}\right)\right\},\eqn\AMP\
$$
where  ${C}_{ffZ}=g_1^2
g_2^2({{\hat g}_{L1}^{q~2}}{{\hat g}_{L2}^{q~2}}+
 {{\hat g}_{R1}^{q~2}}{{\hat g}_{R2}^{q~2}})$;
$\hat s, \hat t,$ and $\hat u$ are the Mandelstam variables;
 $\tilde{u}=\hat{u}/\hat s $,
$\tilde{t}=\hat{t}/ \hat s$; $m_{1,2}=M_{1,2}^2/\hat{s}$;
$m_\pm=m_1\pm m_2$;
 and $M_{1,2}$ are
the respective masses for $Z,\ Z'$.
The total quark cross section is:
$$
\sigma=\int_{\hat{t}_{min}}^{\hat{t}_{max}}
{{d\sigma}\over{d\hat{t}}}\, d\hat{t} =
{C\over{4\pi\hat{s}}}\left\{ {{1+m_+^2}\over{1-m_+}}\ln{{1-m_+ +\beta}
\over {1-m_+-\beta}}-2\beta\right\},
\eqn\EQDKR$$
where $\hat{t}_{\{min,\, max\}}= {1\over2}\hat{s}\{m_+-1\pm\beta\}$,
and $\beta=(1-2m_++m_-^2)^{1/2}$.

In $U_1'$ gauge theories one has $[Q',T_i]=0$. Then, neglecting mixing,
the same formulae \AMP\ and \EQDKR\  apply to
$q_1{\bar q_2}\rightarrow Z'W$,
 with the appropriate
replacements of the gauge couplings and the $W$ mass.
Here, $Q'$ is the
generator of $U_1'$ and the $T_i$'s are the $SU_{2L}$ generators.

On the other hand, in
 theories with mixing the effective $\ghat{L2}{q_1}\not=
\ghat{L2}{q_2}$.  The $\hat{t}$ and $\hat{u}$ channel fermion exchange
contributions become more complicated, and there is also another
contribution from the
non-Abelian graph (see Fig. 1c)
due to  the $Z'Z$ mixing $\propto \theta_{Z'Z}$,
which is responsible for the restoration of unitarity.
 Thus, formulae \AMP\  and \EQDKR\ are not
strictly correct  in this case. However, in gauge theories
with $M_2\gg M_1$ and small $Z-Z'$
mixing  (as constrained from
experiments)\Ref\footVII{
See Refs. \rfii-\rfv, and P.~Langacker, Phys. Rev. {\bf D30}, 2008
(1984).}
$(\ghat{L2}{q_1}-\ghat{L2}{q_2})\sim  \theta_{Z'Z}
\propto (\ghat{L2}{q})\times M_1^2/M_2^2$.
 Note also that
$M_1^2\ll M_2^2\le
\hat s$ is the limit  in which unitarity
should be restored.
 It can   then be shown, using related formulae to those\refmark{\BSM}
for $ZW$ production in the
standard model, that
\AMP\ is still valid in the leading
order $ (\ghat{L2}{q_1}\sim \ghat{L2}{q_2})=
(\ghat{L2}{q})$; the
 correction is suppressed by a factor of
$(\theta_{Z'Z} /\ghat{L2}{q})^2 \times M_2^2/M_1^2\propto
M_1^2/M_2^2$, and is thus negligible. However, in
compositeness-motivated theories (\eg, model {\it (vi)}  with
$Z''$) the gauge
invariance need not be restored;
in this case the analogs \AMP\ and \EQDKR\
violate unitarity.
In such models the
rare decays $Z'\to f_1\bar f_2 W$
also have anomalous rates
in
violation of unitarity.\refmark{\rfxiv}

In models with   heavy charged gauge bosons
$W'$  the production cross sections for the  processes
$q\bar q \to W'W$ can be
evaluated using formulas  \AMP\ and \EQDKR, with the corresponding
values of the gauge couplings and masses, provided one neglects $WW'$
mixing. Since $q\bar q\to W'W$
  directly probes the $({\hat g^f_{L2})}^2$,
it should be strongly
suppressed (by the square of the
$W-W^\prime$  mixing angle
 or by $m_f^2/M_{W^\prime}^2$)
in the  left-right symmetric models. Absence of such events would
thus
provide a good check that the coupling of $W^\prime$ is right-handed.
This is a   second independent check of the right-handedness
 of $W'$, the first  being the absence of
 rare decays $pp\to W'\to f\bar f W$.\refmark{\rfxiv}

The second process with a heavy charged $W'$ involves the production
cross section for $q_i{\bar q_j}\to W'Z$.  In this process the
non-Abelian vertex (see Fig. 1d) $W'W'Z$
is    important even in the absence of mixing,
\ie , it is not suppressed by
$M_1^2/M_2^2$.  In the left-right symmetric models
the squared amplitude  for this process is of
the form\rlap:\refmark\BSM
$$\eqalign{
{\cal M}_{W'Z}=16\pi{{d\sigma}\over{d\hat{t}}}=&2{{g_1^2g_R^2
(\hat{g}^q_{R2})^2}\over\hat{s}^2}     \{
\left({\rho_{W'Z}\over{1-m_2}}
\right)^2 A +{{2\rho_{W'Z}}\over{1-m_2}}
\left[-\hat{g}^i_{R1}I(t)+
\hat{g}^j_{R1}I(u)\right]+\cr
&\left(
\hat{g}^i_{R1}-\hat{g}^j_{R1}\right)^2E +(\hat{g}^i_{R1})^2{\gamma
\over\tilde{u}^2}+ 2\hat{g}^i_{R1}\hat{g}^j_{R1}{m_+\over{\tilde{u}
\tilde{t}}} + (\hat{g}^j_{R1})^2 {\gamma\over\tilde{t}^2}
\} }
\eqn\EQMWPZ$$
with
$$\eqalign{A=& {1\over{m_1m_2}}
\left[ {\gamma\over4} \left( 1-2m_+ +3m_+^2 -2m_-^2\right)+
{m_+\over2}\left(1-2m_++m_-^2\right) \right],\cr
I(t)=& {1\over{m_1m_2}} \left[ {\gamma\over4}
\left( 1-m_+-{{m_+^2-m_-^2} \over\tilde{t}} \right) +{m_+\over2}
\left( 1-m_+ + {{m_+^2-m_-^2}\over{2 \tilde{t}}}\right)  \right],\cr
E=& {1\over{m_1m_2}} \left({\gamma\over4}+{m_+\over2}\right)}
\eqn\EQAIE$$
Here $\gamma=\tilde{t}\tilde{u}-m_1m_2$. The quantities
 $\tilde{u}$, $\tilde{t}$,
$m_{1,2}$,  and
$m_\pm$ are defined after Eq.~(3.1) with $M_1=M_Z$ and
$M_2=M_{W'}$.
The couplings $\hat{g}_{R1}^{i,j}$ and $\hat{g}^q_{R2}$ are the
corresponding charges for $ (i,j) $
quark couplings to $Z$ and $W'$, respectively,
while $g_1 \rho_{W'Z}$ corresponds to the strength of the
$W'W'Z$ non-Abelian vertex.
In the left-right symmetric model one expects
$\rho_{W'Z} = \mp \sin^2 \theta_W$ for $W'^{\pm}$, respectively.
Note, that this process not only provides a way to
measure the strength of the  right-handed $W'$ coupling
but is also a test  of  the non-Abelian vertex $W'W'Z$.

The total production
cross section $\sigma_{V'V}$  for  the above processes
 is obtained in a straightforward manner using the
quark distribution functions\REF\EHLQ{ E. Eichten,
I. Hinchliffe, K. Lane and C. Quigg,
Rev. Mod. Phys. {\bf 56}, 579 (1984).} of Ref.~\EHLQ.
This cross section is compared
 with the basic process $ pp\to Z^\prime\rightarrow \ell^+ \ell^-$
by  defining the ratios
$$R_{Z'V}={{\sigma
(pp\to Z'V)B(Z'\to \ell^+\ell^-)}\over{
\sigma (pp\to Z')B(Z'\to \ell^+\ell^-)}}, \eqn\ratzpv$$
 with $V=Z$ or $W$ decaying into leptons or quarks.
We define the cross section for $pp\to Z'W$ as
the sum over $W^+$ and $W^-$.
 In the models with heavy charged
gauge bosons the ratios
$$R_{W'V}={{\sigma (pp\to W'V)B(W'\to \ell{\bar n_\ell}+
{\bar\ell} n_\ell)}\over{\sigma (pp\to W')B(W'\to \ell{\bar n_\ell}
+{\bar\ell} n_\ell)}}\eqn\ratwpv$$
can be defined   analogously.
The branching ratios involve decay modes with
charged leptons, which provide  clean signals, especially for
$Z'$.

In Fig. 2a and 2b
 we plot the ratios $R_{Z'W}$ and $R_{Z'V}$ as a function
of the $Z'$ mass along with  typical statistical
error bars for a
one year ($10^7$s) run
at  the LHC (projected luminosity
 $10^{34} \hbox{cm}^{-2}\hbox{s}^{-1} $). The error bars are estimated
using the total number of $Z'V$ events shown in Figs. 3a and 3b.  These
assume the specific branching ratios described above, and are presented
only for illustration.  The $R$ values themselves are independent of
the number of exotic decay channels.  The production cross sections
$pp\rightarrow V'$ are presented in Refs.~\LRR-\HRII.
 The   $R$'s increase with the $V'$ mass, but the
statistical error bars are  too large for
$M_{Z'}\ge 2$ TeV for the measurement to be useful. For the projected
SSC luminosity of $10^{33}\hbox{cm}^{-2}\hbox{s}^{-1}$ there are
roughly half as many $V'V$ events for $M_{Z'}\le1$ TeV and equal
numbers for $M_{Z'}>2$ TeV.  The $R$ values are typically 20\% larger
for the SSC.
This is an example  in which increased luminosity would
significantly improve the statistics, and
thus allow one to study the properties of new gauge bosons in a higher
mass range.

In the following we study the sensitivity of the
$R$'s for  different models with $M_{Z'}=1$ TeV at the LHC.
In  Table II the values of $R_{Z'W}$ and  $R_{Z'Z}$ are given
for different models
along with their statistical error bars.
The formula (3.1) does not apply to the $Z''$ for $R_{Z'W}$
because $\ghat{L2}{u}\not=\ghat{L2}{d}$; however Eq. \EQMWPZ\ can be used
with appropriate changes for the masses and gauge couplings, without
the contribution of the non-Abelian graph. Clearly, such an
amplitude badly violates unitarity.
For comparison the values of the ratio $r_{\ell \nu W} =
B(Z'\to \ell\nu W)
/B(Z'\to \ell^+\ell^-)$ of the gold-plated rare decay $Z'\to \ell\nu W$
(see  Fig. 4 for the corresponding Feynman diagram),
 as well as the forward-backward asymmetry $A_{FB}$, are given
along with
their statistical error bars.
The statistical errors  for the $R$'s are
slightly larger than those of $r_{\ell\nu  W} $
and the forward-backward asymmetry, but are still sufficiently small
for $M_{Z'}=1 $ TeV and the
projected LHC
luminosity. In the case of   $Z_{\psi}$ and
$Z_{\eta}$ the error bars
are too large to distinguish between the
two models.

 In Fig.~5     the $ R_{Z'V}$'s
 are plotted
 as a function of
 $\cos\beta$ for the model {\it (iv)} with $Z(\beta)$,
  and the estimated
 error bars are displayed.  Again, the $R_{Z'V}
$'s are a sensitive probe of these models, although there are
 clearly ambiguities.

In Fig. 6  the $R_{Z'V}$'s  are plotted
as a function of $\kappa$ for the left-right models {\it (v)}
along with  typical error bars.
  The
$R_{Z'V}$'s  are   sensitive functions of $\kappa$, especially in the
 interesting region
 $\kappa<1.1$, and could provide a measurement of $\kappa$
independent of measurements from $r_{\ell\nu W}$ and $A_{FB}$.

In Fig.~7 is plotted the predictions for $R_{Z'W}$ vs. $R_{Z'Z}$ for
the $\chi, \psi, \eta$ and $LR\ (\kappa = 1)$
models, along with
expected error bars for $M_{Z'}=1$ TeV.

 In Fig.~8  the ratio  $R_{W'Z}$ (defined in \ratwpv )
is shown for $W'$ in the left-right  symmetric model as a function of
$M_{W'}$  for the LHC and SSC.  This is an absolute prediction
of the model and is a sensitive test of both the right-handed $W'$
coupling (independent of the test from the nonobservation of
$pp\rightarrow W'W$), but also of the non-Abelian $W'W'Z$ vertex.

We  now discuss backgrounds.
 $Z'V$ production,  with $Z'$  subsequently decaying
into  charged leptons and $V$ into hadrons,
charged leptons, or missing neutrinos,
 are clean events without  major   standard model
backgrounds.
$W'V$ productions with   $W'$ decaying
to $\ell  n_\ell$ would be very clean if the $n_\ell$ is heavy and
decays in the detector.  Otherwise, there could in principle be a
standard model
background  from $pp\to WV$ with $W\to \ell\nu_\ell $. However, this
background can be cleanly eliminated  at a loss of only few percent
of the signal by requiring the transverse mass $m_{T\ell n_\ell}$
 of  the  $\ell n_\ell$ system to be larger than  $90$ GeV.

The calculations here utilize the lowest-order expressions in Eqs.
 (3.1), (3.2) and \EQMWPZ\  for
the cross section and the quark-distribution functions in Ref.~\EHLQ.
Of course, if a heavy $V'$ were actually observed, the calculations
would have to be redone using up-to-date distributions and including
${\cal O}(\alpha_s)$ QCD corrections.

\chapter{ Extracting Information From Experimental Data}

We have three types  of independent
experimental signals:
(i) the forward-backward asymmetry $A_{FB}$,\refmark{\LRR}
(ii) the rare decays
$pp\rightarrow V'\rightarrow f_1{\bar f_2}V$,\refmark{\rfxiiip,\rfxiv}
and (iii) the production
cross section $pp\rightarrow V'V$, which
provide useful complementary
probes to extract information about the nature of the gauge couplings.
When expressed in terms of appropriate ratios all are independent of
uncertainties from $\Gamma_{tot}$ and the overall coupling strength
$g_2$.

In this chapter we present a systematic and model-independent
approach for
extracting  information concerning the gauge couplings of the quarks
and leptons from  the above
experimental signals.  We will generally assume that the couplings of
the new $Z'$ are family universal, that the extended gauge group
commutes with the standard model (\ie, $[Q',T_{iL}]=0$), and that
$Z-Z'$ mixing can be ignored.  However, we will first comment on the
possibility of actually testing these assumptions.

\noindent{\sl~~~Interfamily Universality.}
The signals that would most
clearly test the $e-\mu$
family universality are the branching ratios for
$B(Z'\rightarrow e^+e^-)$
versus $B(Z'\rightarrow\mu^+\mu^-)$.  In order to
test the third-family universality,
the measurement of the branching ratio
$B(Z'\rightarrow\tau^+\tau^-)$ is needed.
In spite of having missing neutrinos
in the dominant decay mode $\tau\rightarrow\pi\nu$, it might be
possible\Ref\rfAAC{J. Anderson, M. Austern, and R. Cahn, LBL preprint
LBL--31858, February (1992).} to make such measurements.
The branching ratios
would allow one to determine the interfamily universality
in the lepton sector for the quantities $\ghat{L2}{\ell~2} +
\ghat{R2}{\ell~2}$.  Other combinations of $\ghat{L2}{~2}
$
and $\ghat{R2}{~2}
$
would be probed by comparing $A_{FB}^\ell$ and $r_{\ell\nu W}$ for
$\ell=e$ and $\mu$.

\noindent{\sl~~~ Nature of Enhanced Gauge Structure.} The most likely
possibility for a new $Z'$ in an extended gauge group is that it
corresponds to a $U_1'$ or other group which commutes with the
standard model, $[Q',T_i]=0$.  That can in principle be
tested\refmark{\rfxiv,\HR} (in the absence of $Z-Z'$ mixing) from the
ratio $r_{\ell\nu W}/r_{\nu\nu Z}$, where $r_{\nu\nu Z}=B(Z'\rightarrow
Z\nu\nu)/B(Z'\rightarrow\ell\ell)$, which depends only on the $Z'$ mass
provided $[Q',T_i]=0$ (see below).  In enlarged gauge groups with
$[Q',T_i]\not=0$ there are additional non-Abelian contributions, while
for non-gauge theories (\eg, composite vectors), $r_{\ell\nu W}$ and
also $R_{Z'W}$ may have anomalous or senseless values.  Unfortunately,
as will be discussed below
$r_{\nu\nu Z}$ suffers from a large standard model background
from $pp \rightarrow ZZ$
\refmark{\rfxiv,\PACO}  and is thus probably
 unmeasurable at hadron
colliders.

\noindent{\sl~~~$Z-Z'$ Mixing.} As discussed in Chapter~3 small $Z-Z'$
mixing angles will have a negligible effect on the $pp\rightarrow V'V$
rate.  One place where mixing could be relevant for colliders is
in the decay
 $Z'    \rightarrow WW     $\refmark{\rfx -\rfKaYu}.
Although the rate is
suppressed by the   square of the  mixing angle
($\propto C\left(M_1/M_2\right)^4$), the
longitudinal components of the gauge bosons give an  enhancement
 $\propto
\left(M_2/M_1\right)^4$. A precise measurement of $C$ would yield
valuable information on the symmetry breaking sector of the
theory\refmark{\rfi,\rfix}.  
However,
leptonic modes
  suffer from a serious standard model background
$pp\to WW$.
The semi-leptonic  modes have additional QCD
backgrounds from   $pp\to W+jets$\rlap,\refmark{\rfxii} although it may
still be possible to observe the decay with appropriate
cuts\rlap.\refmark\rfKaYu\


Another possibility\refmark\HR\ is to search for $Z'\rightarrow
WW\rightarrow W\ell\nu$ indirectly by looking for an anomalous ratio
$r_{\ell\nu W}/r_{\nu\nu Z}$ compared to that predicted from the
diagrams in Fig.~4.   However, that is not practical due to the
difficulty of observing\refmark{\rfxiv,\PACO} $r_{\nu\nu Z}$.

A better approach would be
 to try to separate the $W\ell\nu$ events due to mixing
from those from rare decays directly, since they have
very distinct kinematic
distributions. In the $Z'$ rest frame a
 majority of the rare decay events are back-to-back fermions with
invariant mass close to $M_{Z'}$ and a soft $W$ emitted, while the
mixing events
consist of back-to-back $W$'s in the $Z'$ rest frame.
With  the second $W$ decaying hadronically
one can determine
the  transverse mass $m_{T\ell\nu}$ of the
$\ell\nu$ in the $pp$ center of mass. $m_{T\ell\nu}$
must be $<M_W$ for the
mixing events, while some $96\%$ of the rare decays of a 1 TeV $Z'$
have $m_{T\ell\nu}>90$ GeV at the LHC.\refmark{\rfxiv} Thus,
the kinematic
regions  of the two types
of events  are essentially disjoint.
However, both processes have a major background from
$pp\rightarrow WW$. As described below, a
cut $m_{T\ell\nu}\ge 90$ GeV  removes
this background
to the rare decays $r_{\ell\nu W}$, but at the same time it
eliminates
the events due to mixing.  The best prospect for probing $Z-Z'$ mixing
is therefore
in future precision experiments\refmark\rfv\ rather than hadron
colliders.

\noindent{\sl~~~Nature of Gauge Couplings for Leptons and Quarks:}
In the following we study the
extraction of   information on the gauge couplings from the
three types of the
proposed signals: rare decays,
 forward-backward asymmetry,  and $V'V$ production.
Assuming family universality, $[Q',T_i]=0$, and neglecting $Z-Z'$
mixing, the relevant quantities to distinguish different theories are
the charges, $\hat{g}^u_{L2}=\hat{g}^d_{L2}\equiv\hat{g}^q_{L2}$,
$\hat{g}^u_{R2}$, $\hat{g}^d_{R2}$, $\hat{g}^\nu_{L2}=\hat{g}^e_{L2}
\equiv\hat{g}^\ell_{L2}$, and $\hat{g}^\ell_{R2}$, and the gauge
coupling strength $g_2$.  The overall scale of the charges (and $g_2$)
depends on the normalization convention for $\Tr Q'$\rlap,\refmark\rfii\
but the ratios characterize particular theories.  There is little
chance of  ever determining the signs of the charges at hadron
colliders (some information is possible from precision experiments, as
is mentioned below), so we will concentrate on the four ``normalized''
observables,
\def\denom{{(\hat{g}^\ell_{L2})^2+(\hat{g}^\ell_{R2})^2}}
$$\gamma_L^\ell\equiv{{(\hat{g}^\ell_{L2})^2}\over\denom}\eqn\gamll$$
$$\gamma_L^q\equiv{{(\hat{g}^q_{L2})^2}\over\denom}\eqn\gamlq$$
$$\gamma_R^u\equiv{{(\hat{g}^u_{R2})^2}\over\denom}\eqn\gamru$$
$$\gamma_R^d\equiv{{(\hat{g}^d_{R2})^2}\over\denom}.\eqn\gamrd$$
It will also be convenient to introduce
$$\tilde{U}\equiv\gamma_R^u/\gamma_L^q\eqn\tilu$$
$$\tilde{D}\equiv\gamma_R^d/\gamma_L^q.\eqn\tild$$
The values of $\gamma_L^\ell$, $\gamma_L^q$, $\gamma_R^u$,
$\gamma_R^d$, $\tilde{U}$, and $\tilde{D}$ for the $\chi$, $\psi$,
$\eta$, and $LR$ models are listed in Table~III.  It is seen that they
vary significantly from model to model.  In principle, all of the
$\gamma$'s can be determined at hadron colliders.  In practice,
however, only $\gamma_L^\ell$ and $2\tilde{U}+\tilde{D}$ can be well
determined for projected LHC and SSC luminosities.

For rare decays   one observes
the ratios\refmark\rfxiv\ $r_{f_1 f_2V}\equiv B(Z'\rightarrow f_1 {\bar
f_2} V)/B(Z'\rightarrow\ell^+\ell^-)$, where $V=W$ or $Z$.  We always
sum over $\ell=e,\mu$; over $W^+$, $W^-$; and over the neutrino
flavors.
 The expressions for
 $r_{f_1f_2(W,Z)}/a_{W,Z}$
 are bounded\Ref\rfnti{
 The expressions must of course be modified to account for kinematic
cuts.} to a specific range.\refmark{\rfxiv,\HR}
 Here
$$a_{W,Z}\simeq{\alpha\over{6\pi \cos\theta_W^2\sin\theta
_W^2}}
\left[ \ln^2\mu +3\ln\mu
+5 -{\pi^2\over3}\right]\eqn\awz
$$
 are kinematic factors which only depend on
 $\mu \equiv M^2_{Z,W}/M^2_{Z'}$.
For $M_{Z'}=1$ TeV, $a_Z$ ($a_W$) $=0.068$ ($0.080$). For example,
$${r_{ffZ}\over a_Z}={{(\hat{g}^f_{L1})^2 (\hat{g}^f_{L2})^2 +
(\hat{g}_{R1}^f)^2 (\hat{g}^f_{R2})^2}\over{(\hat{g}^{\ell}_{L2})^2 +
(\hat{g}^{\ell}_{R2})^2}}.\eqn\rffza$$
One can
write the leptonic $r$'s as
$${r_{\ell\ell Z}\over a_Z}=0.0529\pm0.020\gamma_L^\ell,\ \
{r_{\nu\nu Z}\over a_Z}=0.375\gamma_L^\ell\le0.375,\ \
{r_{\ell\nu W}\over a_W}=0.77\gamma_L^\ell\le0.77,\eqn\EQLNW$$
and the hadronic $r$'s as
$${r_{{had}\, Z}\over a_Z}=1.17\gamma_L^q+0.071\gamma_R^u
+0.026\gamma_R^d,\ \
{r_{{had}\, W}\over a_W}=2.31\gamma_L^q,\eqn\EQHADW$$
where we have used $\sin^2\theta_W\simeq0.23$. (We define $r_{had\,
W,Z}$ as decays into light hadrons $+(W,Z)$ only; \ie, those including
the $t$-quark are not included.)
Since $\gamma_L^\ell$
 is in the range  $(0,1)$, the
expressions  \EQLNW\ are limited to
 a certain range.
Also, one has $(r_{\ell\nu W}/r_{\nu\nu Z})\times
( a_Z/a_W) =2.05$  (so that
all models satisfying
$[Q',T_i]=0$ lie on a straight line\refmark{\rfxiv,\HR}
 $r_{\nu\nu Z}$ vs. $r_{\ell\nu W}$
 for a fixed value of $M_{Z'}$, and $(r_{had Z}/r_{had W})\times
(a_W/a_Z)\ge0.51$.  These predictions follow directly from $[Q',T_i]=0$,
small $Z-Z'$ mixing, and family universality, and any deviation would
imply a breakdown of one of these assumptions.
The $r's$ are shown for the models considered in Ref. \rfxiv, \rfix.

The quantities in (4.9) and (4.10) in principle determine
$\gamma_L^\ell$, $\gamma_L^q$, and $\gamma_R^u+{3\over8}\gamma_R^d$.
In practice, however, the dependence on the $\gamma_R$'s is very weak.

In hadron colliders the
hadronic ratios $r_{had\,Z}$ and $r_{had\,W}$
suffer\refmark\rfxiv
 from serious QCD  backgrounds. Similarly,
$r_{\nu\nu Z}$ suffers from the standard model background
$pp\to ZZ$, with one $Z\to \nu\bar \nu$.
Even with the second $Z$ fully reconstructed by its hadronic or charged
lepton decays, the only observables are the energy $E_Z$ and angle
$\cos\theta_Z$ in the $pp$ center of mass.  For a 1 TeV $Z_\chi$, for
example, the signal/background is $\simeq10^{-2}$, as shown in Fig.~9a.
The rare decay events have a somewhat harder spectrum, and a flatter
$\cos\theta_Z$ distribution for large $E_Z$, but even by applying
severe cuts (\eg, $\cos\theta_Z<0.6$, $E_Z>300$
GeV), the S/B ratio is still only $10^{-1}$, and one has lost most
of the signal. Hence $r_{\nu \nu Z}$ is probably unobservable
at hadron colliders.  However, $r_{\nu\nu Z}$ and
$r_{had\,W,Z}$
might be useful at future $e^+e^-$  machines.

On the other hand,
$r_{\ell\ell Z}$ provides a very clean signal. Unfortunately,
it has a weak dependence\refmark{\rfxiv}
 on $\gamma_L^\ell$ due to the fact that $|\hat{g}^\ell_{L1}|\simeq
|\hat{g}^\ell_{R1}|$ for $\sin\theta_W^2
\simeq0.23$,
and thus $r_{\ell\ell Z}$ serves only as a consistency check.

 The backgrounds for the ratio $r_{\ell\nu W}$
 can be
eliminated with appropriate cuts.
As discussed earlier, the mode with $W$ decaying hadronically
can be separated from  the standard model
 background  from $pp\rightarrow WW$   (as well as the
events due to $Z-Z'$ mixing)
by requiring\refmark\rfxiv\ the transverse mass $m_{T\ell\nu}\ge
90$ GeV. This cut reduces
the  signal  only by $4\%$ for $M_{Z'}=1$ TeV at the LHC (Fig.~9b).
Events with $W\rightarrow\ell\nu$ may also be observable with
appropriate cuts\rlap.\refmark\PACO\
The ratio  $r_{\ell\nu W}$ thus provides
 the gold-plated signal
  which tests the left-handed coupling of leptons $\gamma_L^\ell$.
One expects an accuracy of (2-10)\% on $\gamma^\ell_L$ at the LHC
for the specific
models considered here.

The second probe is the forward-backward asymmetry. For $pp\rightarrow
Z'\rightarrow\ell^+\ell^-$,
$$A_{FB}={{\left[\int_0^{y_{max}}-\int_{y_{min}}^0\right]\left[ F(y)-B(y)
\right]\,dy}\over{\int_{y_{min}}^{y_{max}}\left[ F(y)+B(y)\right]\,dy}},
\eqn\EQAFB$$
where $F(y)\pm B(y)=[\int_0^1\pm\int_{-1}^0] \,d\cos\theta (d^2\sigma/
dy\, d\cos\theta)$ and $\theta$ is the $\ell^-$ angle in the $Z'$ rest
frame.  $A_{FB}$ can be
be expressed in terms of gauge
 couplings  as:
$$A_{FB}={3\over4}\left(2\gamma_L^\ell-1\right) 0.58{{1
-0.75\tilde{U}-0.25\tilde{D}}\over{1
+0.68\tilde{U}+0.32\tilde{D}}}\eqn\EQAFBP$$
for a 1 TeV $Z'$ at the LHC.
Here the quark distribution functions of Ref.~\EHLQ\ were used.
$A_{FB}$   thus    depends on $\gamma_L^\ell$,
and on the combination $\simeq2\tilde{U}+\tilde{D}$ of right-handed
currents. $A_{FB}$ is shown as a function of $M_{Z'}$ for various
models in Fig.~10, along with the analogous $A_{FB}$ for heavy $W^\pm_R$
production.  It is seen that the statistical errors are small enough to
discriminate effectively up to $\simeq2$ TeV, provided that good enough
lepton charge identification can be achieved.
$A_{FB}$ is shown as a function of $\cos \beta$ for model {\it (iv)}
in Fig.~11, and as a function of $\kappa$ for the LR model {\it (v)}
in Ref. \rfix.
Typically, one should be
able to determine $2\tilde{U}+\tilde{D}$ to (5 -- 10)\% for
$M_{Z'}\simeq1$~TeV if $\gamma_L^\ell$ is known independently from
$r_{\ell\nu W}$.

The exact coefficients of $\tilde{U}$ and $\tilde{D}$ in $A_{FB}$
depend on $M_{Z'}$, $E_{pp}$, and the kinematic cuts.  In principle one
could separate $\tilde{U}$ and $\tilde{D}$ by considering $A_{FB}$ for
various ranges of $y$.  However, we have found that this provides
little additional information for the projected LHC and SSC
luminosities.

The forward-backward asymmetry for $pp\rightarrow
W^{\prime\pm}\rightarrow\ell^\pm   n_\ell$ does not distinguish $V+A$
couplings of the $W'$ from $V-A$.  As described above, this information
can be obtained from $pp\rightarrow W'Z$ and the nonobservation of
$pp\rightarrow W'_R W$ and $W'_R\rightarrow W\ell^+\ell^-$.

The third  probe of $Z'$ charges
are ratios $R$ associated with $Z'V$ production. They are of
the form:
$$\eqalign{
R_{Z'Z}&=10^{-3}{{7.0+0.85\tilde{U}+0.09\tilde{D}}\over
{1+0.68\tilde{U}+0.32\tilde{D}}}\cr
R_{Z'W}&=10^{-3}{{22.2}\over
{1+0.68\tilde{U}+0.32\tilde{D}}}}\eqn\RVPV$$
for a TeV $Z'$ at the LHC.
Again, the quark distribution functions of Ref.~\EHLQ\
 were used.
The numerator in the expression for $R_{Z'Z}$
has a weak dependence  on  $\tilde{U}$ and $\tilde{D}$; this is in part
due to the fact that they are weighted
by the squares of the
gauge couplings of the right-handed quarks to $Z$, which
in the standard model have small values.
 This in turn implies that the ratio $R_{Z'W}/R_{Z'Z}$
 is usually
 close to the  numerical constant  3.2;
however, it  can deviate sizably from this value
in the gauge theory models
with $\gamma_R^{u,d}$ large, \eg ,
in the left-right  symmetric model {\it (v)}.

Since the
ratios $R$  are free of major  standard model and QCD backgrounds,
 the expressions in    \RVPV\  provide
 the first clean probe to yield direct
 information on the couplings of {\it  quarks} to
$Z'$.  Except in models with large $\tilde{U}$ and $\tilde{D}$ the
$R$'s mainly determine the combination $2\tilde{U}+\tilde{D}$,
typically with a precision of $\sim$ (10 -- 20)\%.
This is the same quantity
that is probed by $A_{FB}$ (for a known $\gamma_L^\ell$), but the extra
information provides a welcome consistency check.  Again additional
information to separate $\tilde{U}$ and $\tilde{D}$ could in principle
be obtained by using appropriate $y$ cuts, but the statistics are not
adequate for this for projected luminosities.
In the case of models with heavy charged gauge bosons
the ratio $R_{W'V}$ would also yield  information on the
coupling of $W'$ to the quarks.

One should emphasize that in the case of $A_{FB}$ and the $R$'s
the actual numerical coefficients  in \EQAFB\      and \RVPV\
depend on the quark distribution functions. Thus,
while the experimental detection of these signals is clean,
the actual extraction of the value of couplings crucially
depends on precise knowledge of structure functions and higher-order
QCD effects.  The coefficients would have to be recalculated if a heavy
$V'$ were actually discovered.

We would also like to mention
that it would be useful
if one could measure the branching ratio
 $B(Z'\rightarrow q{\bar q})$.
 Then the ratio
  ${1\over3}
{{B(Z'\rightarrow q{\bar q})}\over{B(Z'\rightarrow\ell^+\ell^-)}}=
2  \gamma_L^q+\gamma_R^u+\gamma_R^d$
(counting all 3 families)
 could be an excellent test of the quark couplings.
 However, the QCD background is too large for
this quantity to be easily
measurable with reasonable resolutions.
Thus, the ratios $R$ and $A_{FB}$ remain as the
only tests to extract information about the quark couplings.

Thus, hadron colliders allow determination of the quantities
$\gamma_L^\ell$ and $\simeq2\tilde{U}+\tilde{D}$ with reasonable
precision for a 1~TeV $Z'$.  From Table~III, we see that these
quantities are a reasonable but not total discriminant between models.
This is further illustrated in Figs.~12 and 13, in which the predictions
for $\gamma_L^\ell$ and $2\tilde{U}+\tilde{D}$ are shown along with
typical statistical error bars for the left-right symmetric and $E_6$
models, as functions of $\kappa$ and $\cos\beta$, respectively.  We see
that the combination of $\gamma^\ell_L$ and $2\tilde{U}+\tilde{D}$
determines $\kappa$ without ambiguity within the interesting range
0.6--1.  There is still a two-fold ambiguity in the $\cos\beta$ value
for $E_6$ models.
The approximate
symmetry around $\cos \beta \sim -0.4$ and $\sim 0.8$ is due to the
vanishing of the couplings of the $SU_5$ 5-plet (10-plet) in
Table 1, and cannot be resolved at a hadron collider
by any means that we are aware of.

For some parameter ranges, the $E_6$ and left-right models cannot be
distinguished, as is expected since left-right models can be embedded
in $E_6$.  This is not a major problem, however, because if a $Z_{LR}$
is discovered directly at a hadron collider, the associated $W'_R$
would almost certainly also be seen.  For most symmetry breaking
structures, the $W'_R$ is lighter than the $Z_{LR}$, and the ratio
$M_{W'_R}/M_{Z_{LR}}$ would be a useful probe\rlap.\refmark\rfix\

At the end we would also like to address
 determination of the
 absolute magnitude of $g_2$, which would
shed light on the
nature of the spontaneous symmetry breaking in a grand unified
theory\refmark\RRref\ or left-right model.\refmark{\rfi,\rfix} As
discussed in the Introduction, for a given theory (\ie, a given set of
$\gamma_{L,R}$'s and a normalization convention for $\Tr(Q')^2$) the
strength of $g_2$ can be determined from the combination
$\sigma\Gamma(Z'\rightarrow\ell^+\ell^-)\equiv\sigma B\Gamma_{tot}$,
while $\Gamma_{tot}$ would then allow an estimate of the fraction of
$Z'$ decays into exotic fermions and superpartners.

Interestingly, the low energy,
atomic parity violation experiments could also
 measure the change in the effective weak charge\refmark{\rfii-\rfv}
 $Q_W$,
$$ \Delta Q_W = -2 (2Z+N) \Delta C_{1u} -2 (Z + 2N) \Delta C_{1d},
\eqn\QW$$ where $Z$ and $N$ refer to the atom and
(when $\theta=0$):
$$ \Delta C_{1i} = 2{{g_2^2 M_1^2}\over{g_1^2 M_2^2}}
\left(\hat{g}^\ell_{L2}-\hat{g}^\ell_{R2}\right)
\left(\hat{g}^q_{L2}+\hat{g}^i_{R2}\right),\eqn\EQPNC$$
for $i = u,d$. This and other precision measurements could
provide an independent way
to gain information about the absolute value of the
coupling. The fractional shift in $Q_W$ in cesium
is shown as a function
of $\cos \beta$ for a 1 TeV $Z'$ in Fig. 14. It is seen that
a precision of a few tenths of a \% would yield a rough determination
of $g_2$, but would not resolve the $\cos \beta $ ambiguities.

\chapter{ Conclusions}
In this paper we have proposed the production cross sections
for $V'V$ with $V'=W',Z'$ and $V=W,Z$ as a
 probe of
the nature of the  gauge couplings of fermions
to heavy gauge bosons at future hadron colliders.
This is the third and complementary probe for the
diagnostic study of such couplings,
the first two being the forward-backward asymmetry\refmark\LRR
 and the  rare decays\rlap.\refmark{\rfxiv,\rfxiiip}
The rates for the production of $V'V$ are
 large for $M_{V'}\sim1\hbox{~TeV}$ and the events  with
$V'$ decaying into charged leptons
 are clean with good diagnostics  for the couplings of
quarks to $V'$.
For $M_{Z'}\sim1\hbox{~TeV}$, the LHC at its projected luminosity would
produce a factor of two more events than the SSC at its
projected luminosity;  the SSC would have more events for
$M_{Z'}\ge3\hbox{~TeV}$. For $Z'$ the events probe the nature of the
$U_1'$ group, while for $W'$ one can test the $V+A$ nature of the
currents and the non-abelian $W'W'Z$ vertex.

In addition, we  have presented a study of extracting
information about the nature of the gauge coupling
from the proposed signals.
We have shown that for models with family universality, $[Q',T_i]=0$,
and small $Z-Z'$ mixing there are four ratios of the squares of gauge
charges that are in principle measurable at hadron colliders.  In
practice only two combinations, $\gamma_L^\ell$ and $2\tilde{U}
+\tilde{D}$, are accessible at projected luminosities, and there are
several complementary measures of these.  They have considerable though
not complete ability to discriminate between models.  Complementary
information from $\sigma B$, $\Gamma_{tot}$, $W'$ production,
$M_{W'}/M_{Z'}$, atomic parity violation, and other precision
experiments should provide additional information to resolve
ambiguities and to probe other aspects of the theory.
 Detection of such
signals, and thus the diagnostic study of such gauge couplings,
would clearly improve with an increased luminosity at the SSC,
which would in turn
provide an important window to learn more about gauge structures
beyond the standard model.

This work was supported by the Department of Energy Grant \#
DE--AC02--76--ERO--3071, the SSC fellowship award (M.C.).
 We would like to thank Jiang Liu  for useful discussions.
\endpage
\input TABLES

\noindent{Table I}

 \begintable
  $ SO_{10} $  & $ SU_{5} $ & $ 2 \sqrt{10} Q_{\chi} $ &
  $ \sqrt{24} Q_{\psi} $ & $ 2 \sqrt{15} Q_{\eta} $ \crthick
16  & 10 $(u,d,\bar{u},e^{+})_{L}$  & $-1$ & $1$ & $-2$ \cr
    & $ 5^{*} (\bar{d},\nu, e^{-})_{L}$  & $3$ & $1$ & $1$ \cr
    & $ 1 \bar{N}_{L}$  & $-5$ & $1$ & $-5$ \cr
10  & 5 $(D,\bar{E^{0}},E^{+})_{L}$  & $2$ & $-2$ & $4$ \cr
    & $ 5^{*} (\bar{D},E^{0},E^{-})_{L}$  & $-2$ & $-2$ & $1$ \cr
1   & $ 1  S^{0}_{L}$  & $0$  & $4$  & $-5$ \endtable
\vskip 0.5in
Couplings of the $Z_{\chi}$, $Z_{\psi}$, and
 $Z_{\eta}$ to a 27-plet of $E_6$. The $SO_{10}$ and $SU_{5}$
 representations are also indicated. The couplings are shown for
 the left-handed ($L$) particles and antiparticles. The couplings of
 the right-handed particles are minus those of the corresponding
 $L$ antiparticles. The $D$ is an exotic $SU_{2}$-singlet quark
 with charge $-1/3$. $(E^{0}, E^{-})_{L,R}$ is an exotic lepton
 doublet with vector $SU_{2}$ couplings. $N$ and $S$ are new Weyl
 neutrinos which may have large Majorana masses.
The $Z_{LR}$ couples to the charge $\sqrt{3/5}
\left[ \beta T_{3R}-1/(2\beta)
T_{B-L} \right] $, where $\beta=\sqrt{\kappa^2\cot^2\theta_W-1}$ and
$\kappa=g_R/g_L$.
\vskip 0.5in

\noindent{Table II}
\vskip 0.5in
\begintable
  & $Z_\chi$ & $Z_\psi$ & $Z_\eta$ & $Z_{LR}$ & $Z^{\prime\prime}$
\crthick
$R_{Z'Z}$ &  $\ 0.0021\pm 0.0002$
& $
 0.0046\pm 0.0008$& $   0.0050\pm 0.0007
 $& $   0.0010\pm 0.0001$& $0.0071 \pm 0.0003
$\cr
$R_{Z'W}$ &$   0.0057\pm 0.0004
$& $   0.012\pm 0.001$&$  0.014\pm 0.001
$&
$   0.0005\pm 0.0001$&  $0.310  \pm 0.002 $\crthick
$r_{\ell\nu W}$ &$   0.055\pm 0.0014
$& $ 0.030\pm 0.002
$&$ 0.012\pm 0.001
$&
$  0.022\pm 0.0008
$&  $ 0.26\pm 0.002
$\crthick
$A_{FB}^{e^+e^-}$ & $-0.134\pm0.007$ & $0.000\pm0.016$ &
  $-0.025\pm0.014$ & $0.100\pm0.006$ & $0.045\pm0.004$ \endtable

Ratios $R_{Z'V}$, $r_{\ell\nu W}$, and $A_{FB}$
with their error bars at the LHC for $M_{Z'}=1 $ TeV
for the models described in the text.
\endpage
\noindent{Table III}
\vskip 0.5in
\begintable
\| $\chi$ & $\psi$ & $\eta$ & $LR$\crthick
$\gamma^\ell_L$ \| 0.9 & 0.5 & 0.2 & 0.36 \nr
$\gamma^q_L$ \| 0.1 & 0.5 & 0.8 & 0.04 \nr
$\gamma_R^u$ \| 0.1 & 0.5 & 0.8 & 1.4 \nr
$\gamma_R^d$ \| 0.9 & 0.5 & 0.2 & 2.6 \nr
$\tilde{U}$ \| 1 & 1 & 1 & 37 \nr
$\tilde{D}$ \| 9 & 1 & 0.25 & 65 \nr
$2\tilde{U}+\tilde{D}$ \| 11 & 3 & 2.3 & 139 \endtable
\vskip0.5in
\noindent{\caps Table.~III} Values of $\gamma_L^\ell$, $\gamma_L^q$,
$\gamma_R^u$, $\gamma_R^d$, $\tilde{U}$, $\tilde{D}$, and
$2\tilde{U}+\tilde{D}$ for the $\chi$, $\psi$, $\eta$, and $LR$
($\kappa=1$) models.

\endpage
 {\bf Figure Captions}

{\bf Figure 1.(a,b)} Lowest order diagrams for $q_i{\bar
q_j}\rightarrow V V'$. {\bf(c)} Additional non-Abelian diagram for
$q{\bar q}\rightarrow Z'W$ induced by $Z-Z'$ mixing. {\bf (d)}
 Non-abelian diagram for $q{\bar q}\rightarrow W'Z$, which is present
 even in the absence of mixing.

 {\bf Figure 2.}
The ratios (a) $R_{Z'W}$ and (b) $R_{Z'Z}$ as a function
of the $Z'$ mass along with typical statistical
error bars for the $Z_\chi$, $Z_{\psi}$, $Z_\eta$, and $Z_{LR}$ for a
one year ($10^7$s) run
at the  LHC with a projected luminosity
 $10^{34} \hbox{cm}^{-2}\hbox{s}^{-1} $.  For the SSC the $R_{V'V}$ are
around 20\% higher.

{\bf Figure 3.}
The numbers of events (a) $N_{Z'W}$ and (b) $N_{Z'Z}$
for ${pp\rightarrow Z'V}$ and
$Z'\rightarrow \ell^+\ell^-$, $\ell = e$ or $\mu$ (summed),
assuming a one-year run at the LHC.

{\bf Figure 4.} Diagrams for $V'\rightarrow Vf_1{\bar f_2}$.

 {\bf Figure 5.}
The ratios $R$
     plotted as a
 function of $\cos\beta$ for the general $E_6$ boson $Z(\beta)=\cos\beta
Z_\chi+\sin\beta Z_\psi$. The error bars are for the LHC with
$M_{Z^\prime}=1 \hbox{~TeV}$.

 {\bf Figure 6.}  The ratios $R$  plotted as a
 function of $\kappa=g_R/g_L$. The error bars are for the LHC  with
$M_{Z^\prime}=1 \hbox{~TeV}$.

{\bf Figure 7.} $R_{Z'W}$ vs. $R_{Z'Z}$ at the LHC for the $\chi$,
$\psi$, $\eta$, and $LR$ models, along with the expected error bars for
$M_{Z'}=1$~TeV.

{\bf Figure 8.} $R_{W' Z}$ as a function of the $W'$ mass for
left-right symmetric models at the LHC and SSC. The corresponding
predictions for $W'^{\pm}$ are very similar.

{\bf Figure 9. (a)} $\cos\theta_Z$ distribution for $pp\rightarrow ZZ$,
with one $Z\rightarrow\nu{\bar\nu}$, in the standard model at the LHC
(solid lines),
compared with the signal for a 1~TeV $Z_\chi\rightarrow Z\nu{\bar\nu}$
(dotted lines). The upper (lower) histograms are for no $E_Z$ cut
($E_Z > $ 300 GeV). {\bf (b)} Transverse mass $m_{T\ell\nu}$ for
$pp\rightarrow W^+W^-$, with one $W\rightarrow\ell\nu$, in the standard
model at the LHC (solid line), compared with the signal for a 1~TeV
$Z_\chi\rightarrow W^\pm\ell^\mp\nu$ (dotted line).

{\bf Figure 10.} $A_{FB}$ is plotted as a function of $M_{Z'}$ at the LHC
for the models described in the text.

{\bf Figure 11.}  $A_{FB}$ plotted as a function of $\cos \beta$
for $M_{Z'}$ = 1 TeV at the LHC.

{\bf Figure 12.}  $\gamma_L^\ell$ and $2\tilde{U}+\tilde{D}$ in
left-right symmetric models {\it (v)} as a function of
$\kappa=g_R/g_L$, along with typical statistical error bars.

{\bf Figure 13.}  $\gamma_L^\ell$ and $2\tilde{U}+\tilde{D}$ in the
general $E_6$ models {\it (iv)} as a function of $\cos\beta$, along
with typical errors.

{\bf Figure 14.}  Fractional change (\%) in the atomic parity violation
parameter $Q_W$ in cesium, for $M_{Z'}$ = 1 TeV and $\lambda_g = 1$.
In the LR model {\it (v)} the change $\sim -0.8$\% depends only
weakly on $\kappa$.

\endpage
\refout\end